\documentclass[prl,twocolumn,showpacs]{revtex4}
\usepackage{bm}
\usepackage{graphicx}
\usepackage{amssymb}
\usepackage{amsmath}
\newcommand{\nix}[1]{}

\begin{document}
\title{Circular $\bm{ac}$ Hall Effect} 

\author
{J.~Karch,$^{1}$ P.~Olbrich,$^1$ M.~Schmalzbauer,$^1$ C.~Zoth,$^1$
C.~Brinsteiner,$^1$ M.~Fehrenbacher,$^1$ U.~Wurstbauer,$^1$
M.~M.~Glazov,$^2$ S.~A.~Tarasenko,$^2$ E.~L.~Ivchenko,$^2$
D.~Weiss,$^1$ J.~Eroms,$^1$ R.~Yakimova,$^3$ S.~Lara-Avila,$^4$
S.~Kubatkin,$^4$ S.~D.~Ganichev$^{1}$}

\affiliation{$^1$ Terahertz Center, University of Regensburg,
93040 Regensburg, Germany}
\affiliation{$^2$ Ioffe Physical-Technical Institute, Russian Academy of Sciences, 
194021 St.~Petersburg, Russia}
\affiliation{$^3$ 
%Department of Physics, Chemistry and Biology, 
Link{\"o}ping University,
S-58183 Link{\"o}ping, Sweden}
\affiliation{$^4$ 
%Department of Microtechnology and Nanoscience, 
Chalmers University of Technology,
S-41296 G{\"o}teborg, Sweden}

\begin{abstract}
We report the observation of the circular $\bm ac$ Hall effect where the current
is solely driven by the crossed $\bm ac$ $\bm E$- and $\bm B$-fields of circularly polarized
radiation. Illuminating an unbiased monolayer sheet of graphene with circularly polarized terahertz radiation at room temperature generates - under oblique incidence - an electric current perpendicular to the plane of incidence, whose sign is reversed by switching the radiation helicity. Alike the classical $\bm dc$ Hall effect, the voltage is caused by crossed $\bm E$- and $\bm B$-fields which are however rotating with the  light's  frequency.
\end{abstract}

%73.50.Pz  Photoconduction and photovoltaic effects
%73.50.-h  Electronic transport phenomena in thin films
%81.05.ue  Graphene 
%72.80.Vp  Electronic transport in graphene 
%78.67.Wj  Optical properties of graphene

\date{\today}

\maketitle

For more than a century, the Hall effect has enabled physicists to gain information
on the electronic properties of matter. In Hall's original experiment~\cite{Hall}, a
clever combination of static magnetic and electric fields allowed to determine
the sign and density of charge carriers, opening the door to a more thorough
understanding of electronic transport in metals and semiconductors.  The circular $ac$ Hall effect (CacHE), in contrast, driven by the crossed $ac$ $\bm E$- and $\bm B$-fields of circularly polarized light,
delivers information on the underlying electron dynamics.
The effect remained so far undiscovered as electromagnetic radiation incident upon low dimensional structures causes all sorts of photocurrents stemming from both contact and band-structure specifics. With respect to the latter the newly discovered graphene~\cite{Bib:Novoselov2004}
%, consisting of a monoatomic layer of graphite, 
is an ideal model system as symmetry prevents other helicity driven photocurrents like the 
circular photogalvanic~\cite{IvchenkoGanichev} or
spin-galvanic effect~\cite{Nature02} to occur.

\begin{figure}[t]
\includegraphics[width=0.8\linewidth]{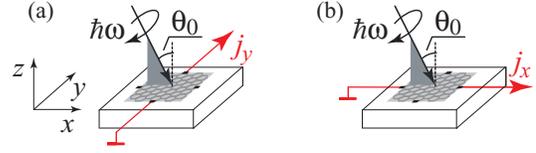}
\caption{Experimental configurations showing the plane of incidence of the radiation,
the arrangement of contacts 
%(black dots) 
at the edges of graphene.
Both (a) transverse and (b) longitudinal arrangements were used to measure the photocurrents.
} \label{fig1}
\end{figure}
Two types of graphene were investigated: large area graphene
prepared by high temperature Si sublimation of semi-insulating
silicon carbide (SiC) substrates~\cite{LaraAvival09} and exfoliated
graphene~\cite{Bib:Novoselov2004} deposited on oxidized silicon
wafers. While both types of samples showed the effect, the micron
sized exfoliated samples displayed an additional edge contribution
(discussed in Ref.~\cite{condmat}) as the spot size of the terahertz (THz) laser of
1~mm$^2$ was larger than the graphene flakes. Hence, we focus on the
large area SiC based samples having areas of $3 \times 3$ and $5
\times 5$~mm$^2$.
Both \textit{n}-
and \textit{p}-type layers were probed. The experimental geometry is
sketched in Fig.~\ref{fig1}. The graphene samples were illuminated
at oblique incidence, where the incidence angle $\theta_0$   was
varied between $-40^\circ$ and +40$^\circ$. The resulting
photocurrent was measured at room temperature for wavelengths
between 90\,$\mu$m and 280\,$\mu$m using either a continuous-wave ($cw$) CH$_3$OH
laser or a high power pulsed NH$_3$ laser~\cite{Ganichevbook}. For
these wavelengths the condition $\omega \tau < 1$ holds, with
$\omega$  the angular frequency of the light and $\tau$ the momentum
relaxation time of electrons (holes) in graphene. The resulting photocurrent is measured
by the voltage drop across a load resistor between pairs of
contacts made at the edges of the graphene square. To prove that the
signal stems from graphene and not, e.g., from the substrate, we
removed the graphene layer from one of the exfoliated samples and
observed that the signal disappeared. The degree of circular
polarization, $P_{\rm circ}= \sin 2 \varphi$, is adjusted by a
quarter-wave plate, where $\varphi$  is the angle between the initial
polarization vector of the laser light and the $c$-axis of the
plate. 
%polarization ellipses are sketched for some angles $\varphi$
%on top of Fig.~\ref{fig2}.

\begin{figure}[t]
\includegraphics[width=0.75\linewidth]{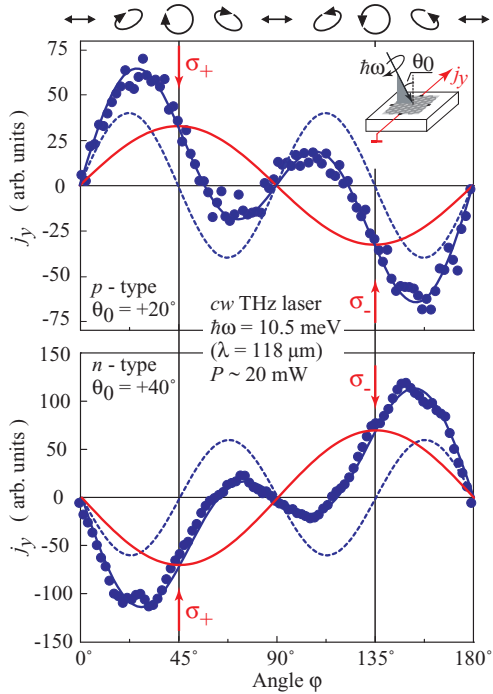}
\caption{Transverse photocurrent $j_{y}$ as a function of the angle $\varphi$
for $p$- and $n$-type graphene.
The ellipses on top illustrate the polarization states for various
$\varphi$.
Full blue lines show fits to the calculated total current $j_A +
j_B$ comprising the circular contribution $j_A$ (CacHE, full red
line) and the linear contribution $j_B$ (dashed blue line).
} \label{fig2}
\end{figure}

The photocurrent for the transversal geometry, $j_y$,  is shown in
Fig.~\ref{fig2} as a function of $\varphi$.
%for \textit{p}- and \textit{n}-type graphene. 
The principal observation made in all
investigated samples is that for circularly polarized light, i.e.
for $\varphi = 45^{\circ}$ and
135$^{\circ}$, the sign of $j_y$ depends on the light's helicity and
the charge carriers' polarity.
% (positive for holes and negative for electrons). 
The overall dependence of $j_y$ on $\varphi$ is more
complex and, at small $\theta_0$,  well described by
\begin{equation} \label{phenom1}
j_{y} = A \theta_0 \sin 2\varphi + B \theta_0 \sin 4\varphi + \xi\:.
\end{equation}
Here, $\xi$  is a polarization independent offset,
ascribed to sample or intensity inhomogeneities.
It does not change with angle $\theta_0$ and is
subtracted from the data of Fig.~\ref{fig2}. The fit parameters $A$
and $B$ describe the strength of the circular contribution $j_A
\propto \sin 2\varphi$ and of the contribution $j_B \propto \sin
4\varphi$ caused by linear polarization. Both contributions
%, $j_A$ and $j_B$ as red curve and $j_B$ as dashed one, 
are shown together with the
resulting fit of the data 
%($j_A+j_B$, solid blue line) 
in Fig.~\ref{fig2}. Note that for purely circularly polarized light,
%i.e. at  $\varphi = 45^{\circ}$ and 135$^{\circ}$, 
the linear contribution $j_B$ vanishes.

\begin{figure}[t]
\includegraphics[width=0.75\linewidth]{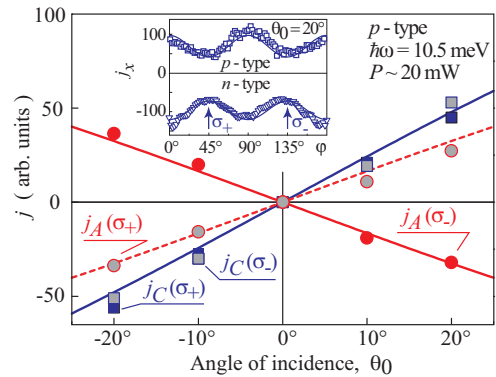}
\caption{Photocurrents $j_A$ (red symbols) and $j_C$ (blue symbols)
induced by circularly polarized light $\sigma_\pm$ ($\varphi=45^\circ$ and
$135^\circ$) as function of the incidence angle $\theta_0$. Open red circles
and blue squares correspond to $\sigma_+$, filled red circles and blue
squares to $\sigma_-$.
The inset shows the $\varphi$ dependence of $j_x$ measured both in
$p$- and $n$-type graphene together with fits according to
Eq.~\eqref{jx}. The solid lines are fits based on Eqs.~\eqref{ja}-\eqref{jc}.
%For all data the 
The constant offsets $\xi$ and $\xi'$ have been
subtracted.
} \label{fig3}
\end{figure}

In the longitudinal geometry [Fig.~\ref{fig1}(b)], only linearly
polarized light gives rise to the $\varphi$-dependence of 
%the current 
$j_x$:
\begin{equation}
\label{jx}
j_x = B \theta_0 (1 + \cos 4\varphi) + C \theta_0 + \xi'.
\end{equation}
This is shown in the inset of Fig.~\ref{fig3} for both \textit{n}- and
\textit{p}-type graphene. A sizable fraction of $j_x$ stems from
the polarization independent contribution $j_C=C\theta_0$,
%proportional to the angle of incidence $\theta_0$. 
whose sign does not reverse with helicity. Both currents
$j_y$ and $j_x$, however, change their signs upon reversing the direction of
incidence 
%; the corresponding dependence  on $\theta_0$ is shown in
(Fig.~\ref{fig3}).

\begin{figure}[t]
\includegraphics[width=\linewidth]{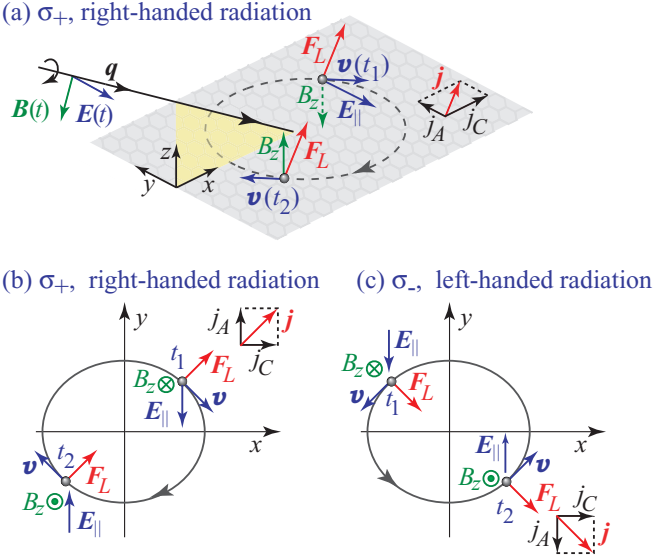}
\caption{ Schematic illustration of the circular $ac$ Hall effect. For
simplicity we assume positive carriers, i.e. holes. (a) 
%Electric and magnetic 
$\bm E$- and $\bm B$-field vectors of $\sigma_+$ polarized light with wave
vector ${\bm q}$ under oblique incidence in the ($xz$) plane. The
solid orbit represents the hole's elliptical trajectory caused by
the $ac$ $\bm E$-field.
The relevant vectors are shown for two instants in time, $t_1$ and
$t_2$, shifted by half a period. ${\bm v_1}$ and ${\bm v_2}$ are the
hole velocities at $t_1$ and $t_2$, respectively, taking retardation
into account. The direction of the Lorentz force $\bm F_L$ due to the $ac$ $\bm B$-field
determines the direction of the Hall current $\bm {j}$.
(b) Top view of (a). (c) Same as (b) but for $\sigma_-$ light.
%polarization.
} \label{figm1}
\end{figure}

The experimental data are well described by the theoretical model,
outlined below. While the longitudinal currents can be explained
along similar lines, we focus on the transverse helicity-driven
current $j_A$.
The basic physics behind the CacHE is
illustrated in Fig.~\ref{figm1}. Here, we consider the classical
regime, where the photon energy is much smaller than the Fermi
energy, $\hbar \omega \ll |E_F| $, fulfilled in the experiment as $|E_F|$ is $\sim
100$~meV while the photon energy $\hbar\omega$ is typically $\sim
10$~meV. For circularly polarized radiation, the electric field
rotates around the wavevector ${\bm q}$, sketched in
Fig.~\ref{figm1}(a) for $\sigma_+$ circularly polarized light. This
leads to an orbital motion of the holes (electrons) illustrated in
Fig.~\ref{figm1}. The CacHE comes into existence due to the combined
action of the rotating electric and magnetic field vectors ${\bm E}$
and ${\bm B}$, respectively. At an instant of time, e.g., at
$t_{1}$, the electron is accelerated by the
in-plane component ${\bm E}_{\parallel }$ of the $ac$ electric
field.
At the same time, the electron with velocity ${\bm v}$ is subjected
to the out-of-plane magnetic field component ${\bm B}_z$.
%perpendicular to the graphene sheet. 
Note, that the velocity ${\bm v}$ does not
instantaneously follow the actual ${\bm E}_{\vert \vert }$-field
direction due to retardation: There is a phase shift equal to
$\arctan(\omega \tau)$ between the electric field and the electron
velocity ${\bm v}$. Only for $\omega \tau \ll 1$
the directions of $\bm v$
%the electron velocity
%and the direction of 
and ${\bm E}_{\parallel }$ coincide.
The effect of retardation, well known in the Drude-Lorentz theory of high frequency conductivity~\cite{Ashcroft},
results in an angle between the velocity ${\bm v}$ and the electric field direction ${\bm E}_{\parallel}$,
which depends on the value of $\omega \tau$.
The resulting Lorentz force $\bm F_L =  e  (\bm v \times \bm B_z)$,
where $e$ is the positive (holes) or negative (electrons) carrier charge,
generates a Hall current $\bm j$,
also shown in Fig.~\ref{figm1}.
Half a period later at $t_2 = t_1 + T/2$, both ${\bm v}$ and ${\bm B}_z$
%the electron velocity and the out-of-plane magnetic field component
get reversed so that the direction of $\bm F_L$
%the Lorentz force 
and, consequently, the current $\bm j$ stay the same. The oscillating
magnitude and direction of ${\bm B}_z$ along the closed trajectory
leads to a periodical modulation of the Lorentz force with non-zero average
causing a non-zero time-averaged Hall current with fixed
direction.

If, as shown in Fig.~\ref{figm1}(c), the light helicity is reversed,
the electric field rotates in the opposite direction and, thus, the %charge
carrier reverses its direction. Hence, the $y$-component of 
$\bm F_L$ at $t_1$ and $t_2$ is inverted. Consequently the polarity of
the transverse, time-averaged Hall current changes. This is the
circular $ac$ Hall effect. On the other hand, we obtain the
longitudinal current $j_x$, which does not change direction when the
helicity flips. This current is also observed in our experiment,
displayed in 
%the inset of 
Fig.~\ref{fig3}. Obviously, flipping the
angle of incidence, $\theta_0 \rightarrow -\theta_0$, results in a
change of the relative sign of  ${\bm E}_{||}$ and ${\bm B}_z$
so that both $j_x$ and $j_y$
%longitudinal and transversal currents 
flip directions.

While the explanation of the CacHE
has been given in a pictorial way above, we resort now to a
microscopic description based on the Boltzmann kinetic equation for
the electron distribution function $f(\bm p, \bm r,t)$, with the
free-carrier momentum $\bm p$, in-plane coordinate $\bm r$, and time
$t$:
\begin{equation}
 \label{kinetic:gen}
\frac{\partial f}{\partial t} + \bm v \frac{\partial f}{\partial \bm r} + e \left(\bm E + \bm v \times \bm B\right)
\frac{\partial f}{\partial \bm p} = Q\{f\}\:.
\end{equation}
Here, $Q\{f\}$ is the collision integral described in terms of
momentum relaxation times $\tau_n$ ($n=1,2\ldots$) for
corresponding angular harmonics of the distribution function. The
electric current density is given by the standard equation $\bm j =
4 e \sum_{\bm p} {\bm v}\, f(\bm p)$, where a factor of 4 accounts
for  spin and valley degeneracies. In order to solve the kinetic
equation \eqref{kinetic:gen}, we expand the solution in powers of
electric and magnetic fields, keeping linear and quadratic terms
only. In the calculation of $f(\bm p)$ and ${\bm j}$, we used the
energy dispersion $\varepsilon_p= \pm v p$ of free carriers in
graphene and the relation $\bm v \equiv \bm v_{\bm p}=v \bm p/|\bm
p|$ between the velocity and the quasi-momentum ($v \approx c/300$,
with $c$ being the speed of light).  Contributions to the
photocurrent appear not only from a combined action of the electric
and magnetic fields of the light wave,
illustrated in Fig.~\ref{figm1}, but also due to the spatial
gradient of the electric field~\cite{perel73}.
As final result we obtain for the helicity driven current
\begin{equation}
\label{ja}
 j_A = A\theta_0 \sin{2\varphi} = q \theta_0 \chi P_{\rm circ}\left( 1 + \frac{\tau_2}{\tau_1}\right) \frac{1 - r}{1 + \omega^2 \tau_2^2},
\end{equation}
flowing in $y$-direction, and the $\varphi$-independent current
\begin{equation}
\label{jc}
 j_C = C\theta_0 =  \frac{q \theta_0\chi}{\omega \tau_1} \left[ 2( 1 + r) + (1 - r) \frac{1 - \omega^2 \tau_1 \tau_2}{1 + \omega^2 \tau_2^2} \right],
\end{equation}
flowing along $x$ (for light propagating in the ($xz$) plane).
Here $q = \omega / c$, 
$q \sin{\theta_0} \approx q \theta_0$, 
$r = d{\rm ln}\tau_1/d{\rm ln}\varepsilon$ and
$\chi = e^3 {\tau_1 (v \tau_1 E)^2 }/[{2 \pi \hbar^2 (1 + \omega^2 \tau_1^2)}]$.

\begin{figure}[t]
\includegraphics[width=0.75\linewidth]{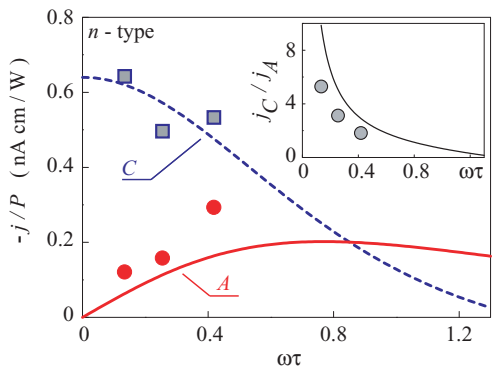}
\caption{Frequency dependence of $A = j_A / \theta_0$ (red dots) and $C = j_C / \theta_0$ (blue squares) as function of $\omega\tau$ for circularly polarized light.
Data are shown for wavelengths between 90~$\mu$m and 280~$\mu$m with the power ranging from 10 kW to 30 kW. The photocurrent $j_C$ is obtained from the current in $x$-direction, which for $\sigma_+$, $\sigma_-$-light reads $j_x=C\theta_0$. The calculated frequency dependence of $j_A$ (Eq.~\eqref{ja}, red solid line) and $j_C$ (Eq.~\eqref{jc}, dashed line) describe the experiment quantitatively well. The inset shows $j_A/j_C$ both for experiment and theory. This plot, independent of the absolute values, shows that the
helicity driven current $j_A$ vanishes for $\omega\tau\ll 1$.
} \label{figm2}
\end{figure}

The results of the calculation are shown 
%by red and blue lines 
in Figs.~\ref{fig3} and \ref{figm2}. The used fitting parameters only
depend on details of the underlying scattering mechanism discussed
below.
Equation~(\ref{kinetic:gen}) provides in addition to $j_A$ and $j_C$ also 
%expressions for 
currents generated by linearly polarized
light,
$j_{B,x} \propto q \theta_0 (1+\cos 4\varphi)$ and
$j_{B,y} \propto q \theta_0 \sin 4\varphi$.
These currents with different angular dependencies are superimposed
on the circular $ac$ Hall effect (and also on $j_C$ discussed above), when
$\varphi$ is varied, and cause a more complex polarization
dependence of the photocurrent (see blue lines in Figs.~\ref{fig2}
and in the inset of Fig.~\ref{fig3}). However, for perfect circularly
polarized light ($\varphi=45^{\circ}$ or $135^{\circ}$), the degree
of linear polarization is zero and the corresponding currents
vanish leaving
the undisturbed CacHE contribution, as shown in Figs.~\ref{fig2} and \ref{fig3}.

%The polarity of the photocurrents displayed in Fig.~\ref{fig2}
%, (a) and (b), 
%and in the inset of Fig.~\ref{fig3}, 
As seen in experiment the polarity of the photocurrents
is opposite for $n$- and $p$-type graphene samples.
This is expected from theory since (i) the $ac$ Hall current $j_y$ as well as the
longitudinal current $j_x$ are proportional to $e^3$ and
(ii) the conduction- and valence-band,
in the massless Dirac model, are symmetric with respect to the Dirac
point. 
In contrast, in typical semiconductors conduction- and valence-band states have different
symmetry properties and the relation between values and polarities
of the 
%transverse 
$ac$ Hall photocurrents is more involved.

Equations (\ref{ja}) and (\ref{jc}) suggest a non-monotonous
frequency dependence of the photocurrents. 
In Fig.~\ref{figm2} the calculated frequency dependence of both $A =
j_A/\theta_0$ and $C=j_C/\theta_0$ are compared quantitatively to
experimental data.
For the momentum scattering time we used the relation $\tau_1 = 2
\tau_2 \propto \varepsilon_p^{-1}$, valid for short range
scattering~\cite{sarma} and relevant for our low mobility samples.
The carrier density $n = 3.8 \times 10^{12}$~cm$^{-2}$ and momentum
scattering time $\tau_1 \approx 2 \times 10^{-14}$~s were obtained
from transport measurements. Apart from the above assumption of
short range scattering, no fit parameter was used. Figure~\ref{figm2}
shows that the theory describes the frequency dependence and the
absolute value of the photocurrent very well. Both
$j_A$ and $j_C$ contribute to the photocurrent for circularly
polarized light. It is remarkable that the helicity driven
current $j_A$ and the polarization independent photocurrent $j_C$ show
completely different frequency dependencies. While $j_C$ does not
change much for $\omega\tau \ll 1$, 
%the helicity driven current 
$j_A$ increases with growing $\omega\tau$ at low frequencies. For large
$\omega \tau $ well above unity both photocurrents decrease
with increasing $\omega$. This property agrees with the model
addressed above 
%(see Fig.~\ref{figm1}). 
The CacHE, i.e. $j_A$, disappears for $\omega\to 0$, since no circular polarization
exists for static fields and the required retardation % for the effect
vanishes. With increasing $\omega$ the retardation becomes important
and the current increases $\propto \omega\tau$. For
$\omega\tau\simeq 1$ the current gets maximal and decreases rapidly
at higher $\omega$, $j_y \propto 1/\omega^4$. In contrast, the
longitudinal current $j_C$ does not depend on the frequency at 
$\omega \tau \ll 1$ and displays its maximum at $\omega\to 0$. The
effect of retardation is just opposite to that on $j_A$: Increasing
$\omega$ reduces the $y$-component of the velocity (Fig.~\ref{figm1}) and hence the
$x$-component of the Lorentz force. As a consequence, $j_C$ drops
with increasing $\omega$, see Fig.~\ref{figm2}. The ratio of $j_C$ and
%the circular current 
$j_A$ is plotted in the inset of
Fig.~\ref{figm2} showing that the role of the circular effect
substantially increases with 
%larger 
$\omega \tau $.
The excellent agreement of theory and experiment shows
that the model covers the essential physics of the circular $ac$ Hall effect.

The photocurrents $j_C$ and $j_A$ are both proportional to the
wavevector $q$ and may, therefore, also be classified as photon drag
effect. In fact, the  polarization independent longitudinal current
$j_C$ is the well-known linear photon drag effect, which was first
treated by Barlow~\cite{Ch7Barlow54} in 1954, observed in bulk
cubic semiconductors~\cite{Ch7Danishevskii70p544,Ch7Gibson70p75} 
and recently discussed for graphene~\cite{condmat,Entin:gr}. 
The circular $ac$ Hall effect, described
here, can be considered as the classical limit ($\omega\tau<1$) of
the circular photon drag effect. 
The latter effect which takes over
at higher frequencies, i.e. for $\omega\tau > 1$, was discussed
phenomenologically~\cite{Ivchenko1980,Belinicher1981} and observed in
GaAs quantum wells in the mid-infrared range~\cite{Shalygin2006}.
In this pure quantum mechanical limit the picture above is inapplicable 
and involves asymmetric optical transitions and relaxation in
a spin polarized non-equilibrium electron gas.
The drag effect in metallic photonic crystals, generating a transverse current due to microscopic voids, was reported recently~\cite{Hatano09}.
%It has also  been reported recently for a lateral metallic
%photonic crystal, but the effect requires modification of local near-surface electromagnetic fields 
%and vanishes in plain films~\cite{Hatano09}.

The appearance of a helicity driven Hall current is
a specific feature of two-dimensional, even centrosymmetric, structures like
graphene.
CacHE is a general phenomenon and should exist in any
low-dimensional system. It is however more readily observable in a
monoatomic layer like graphene, as in multilayered low-dimensional
systems, e.g. quantum wells, the CacHE is masked by the circular
photogalvanic effect~\cite{IvchenkoGanichev}. In monoatomic layers studied here,
however, all photogalvanic effects
vanish, since they require an electronic response to the
out-of-plane component of the electric field. Using graphene
therefore allowed us to identify the circular $ac$ Hall effect
unambiguously, thus providing a novel access to charge carrier
dynamics and scattering.

We thank  J.~Fabian, V.V.~Bel'kov, J. Kamann, and
V. Lechner for fruitful discussions and support.
Support from DFG (SPP~1459 and GRK~1570),
Linkage Grant of IB of BMBF at DLR, RFBR, Russian Ministry of Education and Sciences, President grant for
young scientists and ``Dynasty'' Foundation ICFPM is acknowledged.

\end{document}